# GRAVITATIONAL MICROLENSING, THE DISTANCE SCALE, AND THE AGES

BOHDAN PACZYŃSKI
*Princeton University Observatory*
*124 Peyton Hall, Princeton, NJ 08544-1001, USA*

**Abstract.**
A high optical depth to gravitational microlensing towards the galactic bulge is consistent with current models of the galactic bar. The low optical depth towards the LMC can probably be accounted for by the ordinary stars in our galaxy and in the LMC itself. No conclusive evidence is available yet for the presence or absence of a large number of brown dwarfs or other non-stellar compact objects which might account for the dark matter. There is little doubt that the amount of mass in objects in the range $10^{-8} \leq M/M_\odot \leq 10^6$ will be determined within the next few years with the continuing and expanding searches.

Billions of photometric measurements generated by the microlensing searches have lead to the discovery of $\sim 10^5$ variable stars. In particular, a number of detached eclipsing binaries were discovered in the galactic bulge, in the LMC, and in the globular cluster Omega Centauri. The follow-up observations of these binaries will allow the determination of accurate distances to all these objects, as well as robust age determination of globular clusters.

## 1. Microlensing

Four groups: DUO (Alard 1995), EROS (Aubourg et al. 1993), MACHO (Alcock et al. 1993) and OGLE (Udalski et al. 1993) are conducting the search for events of gravitational microlensing among millions of stars in the Magellanic Clouds (EROS, MACHO), and in the galactic bulge (DUO, MACHO, OGLE). A total of almost 100 events of microlensing by single objects (Alcock et al. 1993, 1995a,c,d, Aubourg et al. 1993, Udalski et al.

1993, 1994a,b,c, Alard 1995, Stubbs et al. 1995), and at least 3 events of microlensing by double object (Udalski 1994d, Alard et al. 1995, Pratt 1995), were reported so far.

## 1.1. PAST AND PRESENT

This is a very young and rapidly developing field, following a science-fiction like suggestion by Paczyński (1986). The practicality of microlensing searches was first envisioned by K. Freeman, D. P. Bennett, and C. Alcock, and the first dedicated camera was built by C. Stubbs, all of the MACHO collaboration. The detection of the first candidate events was reported almost simultaneously by three groups: EROS (Aubourg et al. 1993), MACHO (Alcock et al. 1993), and OGLE (Udalski et al. 1993).

Originally, data processing lagged behind data acquisition, and microlensing events were discovered on computer tapes many months after the recording had been made. The first single event ever recorded was OGLE #10, which peaked on June 29, 1992 (Udalski et al. 1994b). The first double event was OGLE #7 (Udalski et al. 1994d) which had its complicated light variations in June/July 1993. Double lensing events had been estimated by Mao & Paczyński (1991) to make up to 10% of all events.

In the spring of 1994 the real time data processing became possible for the OGLE collaboration, and the first event detected while in progress was OGLE #11 (Udalski et al. 1994c); it peaked on July 5, 1994. Soon after that the MACHO collaboration, which has yearly data rate $\sim 30$ times higher than the OGLE's rate, started first partial, and now full real time data processing. The total number of "real time" events is currently 6 for the OGLE (as announced electronically by Udalski), and $\sim 30$ for the MACHO (Pratt 1995).

Long duration single events should be somewhat time asymmetric because of the acceleration in the Earth's motion around the sun (Gould 1992). This "parallax effect" was first detected by the MACHO group (Alcock et al. 1995d).

The optical depth to microlensing towards the galactic bulge was found to be $\sim 3 \times 10^{-6}$ (Udalski et al. 1994b, Alcock et al. 1995c), while the original theoretical estimates were below $10^{-6}$ (Paczyński 1991, Griest et al. 1991). Theoretical estimates were increased by Kiraga & Paczyński (1994) to somewhat above $10^{-6}$ when the importance of self-lensing by the bulge was recognized, and increased even more when the importance of the galactic bar was noticed (Paczyński et al. 1994). Subsequent calculations, using the best available bar model, gave the optical depth of $\sim 2 \times 10^{-6}$ (Zhao et al. 1995), within one standard deviation of the observational results. According to Zhao et al. (1995) the duration distribution of the OGLE events detected towards the bulge is consistent with no brown dwarfs among the lensing objects, though a 50% contribution by brown dwarfs can be ruled out at only 95% confidence level.

In their original paper the EROS group wrote that their two events towards LMC were consistent with all dark matter in the galactic halo being in the form of MACHOs (Aubourg et al. 1993). The MACHO group, on the basis of many more measurements but only three events, estimated that the MACHOs can account for no more than $\sim 20\%$ of a standard dark halo (Alcock et al. 1995a). Sahu (1994) pointed out that the dominant stellar contribution to microlensing of stars in the bulge of LMC may be due to stars in the bulge of LMC.

Microlensing events are often advertized as achromatic. However, all searches are conducted in very dense stellar fields, where the detection limit is set by crowding. Therefore, many (perhaps most) images are blends, made of two or more stellar images which appear as one. The apparent images are $\sim 1$ second of arc across because of the atmospheric seeing, while the cross-section for gravitational lensing is about $(0.001'')^2$. As a result only one star of a blend is likely to be lensed (DiStefano & Esin 1995). As stars may have various colors, the lensing event may not be strictly achromatic. The contribution of blends was first noticed in the analysis of double lensing events (Udalski et al. 1994d, Alard et al. 1995), but it must be common among all events. It may seem surprising that the the observed events are almost achromatic. There may be two reasons for that. First, in the case of the galactic bulge the detection limit is close to the bulge main sequence turn-off point, and stars have only a small range of colors within a few magnitudes of the turn-off point. Second, some genuine events might have been rejected on the grounds that they appeared to be chromatic.

The calibration of microlensing searches may be done at two levels of sophistication. It may be done at the "catalog level", in which the artificial events are introduced into the database of photometric measurements (Udalski et al. 1994b), or at the "pixel level", in which the artificial events are introduced as artificial stars on the CCD frames (Alcock et al. 1995a, Stubbs et al. 1995). The second method is the truly correct way of doing the analysis, but it is also vastly more time consuming than the first. Fortunately, the search sensitivity as calibrated by the two methods differs by no more than $\sim 20\%$ (Stubbs et al. 1995) as a result of a near cancellation of the two effects: the number of stars subject to microlensing should be increased because of the blending, but the observed amplitudes are reduced by the blending (Udalski et al. 1994b).

The OGLE and MACHO experiments are fairly efficient in detecting events with time scales in the range 10 days – 100 days, but their efficiency falls rapidly outside these limits. The full range of time scales of microlens-ing events which might be detectable with the current technology extends from $\sim$ 10 minutes (Paczyński 1986) to $\sim$ 200 years (Paczyński 1995b). The lower limit is set by the requirement that the stellar disk should be smaller than the Einstein ring to make a significant magnification of the apparent brightness possible. This corresponds to the lens of $\sim 10^{-8}$ $M_\odot$. The upper limit brings us up to $\sim 10^6$ $M_\odot$, the larger masses being detectable by other means (Wambsganss & Paczyński 1992). A sensitive coverage of time scales in the whole range: $\sim$ 10 minutes to $\sim$ 200 may be expected within a few years, but a really high sensitivity is currently limited to the interval 10 – 100 days. The upper limits on the number of very short time scale events, i.e. the optical depth to very low mass objects, currently available from EROS (Aubourg et al. 1995) and MACHO (Stubbs, private communication) are not very stringent.

There is plenty of evidence that almost all reported events are due to microlensing:

1. The light variations are achromatic or almost achromatic, and the light curves are well described by the theoretical formula.
2. The distribution of peak magnifications is consistent with that expected theoretically.
3. The double lensing events are detected at a frequency consistent with the theoretical expectations.
4. The parallax effect has been detected on the longest event, as expected.
5. The the galactic bar has been "rediscovered" with gravitational microlensing (my personal view).

There is no formal consensus on many other issues, as judged from the published papers:

1. What is the reason for the large optical depth towards the galactic bulge (OGLE vs MACHO).
2. What is the optical depth towards the LMC (EROS vs MACHO).
3. Which objects contribute to the LMC microlensing (Sahu vs MACHO).
4. What is the frequency of double lenses (OGLE & DUO vs MACHO).
5. What can be said about the presence or absence of dark compact objects in the galactic halo (EROS vs MACHO vs author).

There is little doubt that within the next few years all these issues will be resolved.

1.2. FUTURE

There is no lack of bold proposals for future microlensing searches from the ground as well as from space (Gould 1995, and references therein). I shall address here just a few topics.

Currently it is possible to estimate the mass of the lensing object only statistically, as the distance to any lens and its transverse velocity are not known. At the same time our knowledge of the mass function for low mass stars and brown dwarfs is very limited. It should be possible to fix both problems with a new observing program: the search for microlensing events caused by the high proper motion stars (Paczyński 1995a). The most difficult step is the detection, in the densest regions of the Milky Way, of a few hundred or a few thousand stars with proper motion in excess of $\sim 0.2$ seconds of arc per year in the magnitude range $18 \leq V \leq 21$, or so. The faint high proper motion stars must be nearby, and their distances can be measured directly and accurately with trigonometric parallaxes (Monet et al. 1992). The same observations will determine the proper motion with accuracy high enough to make reliable predictions about the microlensing events of the distant stars in the Milky Way by the faint high proper motion stars or brown dwarfs. The expected time scale for the events is in the range of 1 – 10 days, i.e. they can easily be followed photometrically. The mass of the lens can be calculated without any ambiguity when the event time scale, as well as the distance and the proper motion of the lens are all measured.

The search for planets (Mao & Paczyński, 1991) will require a major upgrade of the current ground based searches, it may even call for space probes capable of monitoring the events from a distance of $\sim 1$ AU (Gould 1995, and references therein). There is a major problems to solve: how to prove convincingly that a short time scale feature in the light curve is actually caused by a planet, rather than by stellar variability. As the diversity of possible light curves is enormous it is not clear how to conduct statistical tests, like those which can be done for single lensing events.

There is plenty of room for improvement of hardware and software of the ground based searches. On the basis of published OGLE and MACHO results one can find that on average $\sim 20$ pixels on a CCD are needed to measure brightness of a single (possibly blended) stellar image. It is very likely that higher efficiency might be achieved with some "frame subtraction" technique, most unfortunately referred to as "pixel lensing" (Crotts 1992, Colley 1995, Gould 1995).

Let $N$ be the total number of photometric measurements of all stars monitored in a given experiment, $\tau$ the optical depth to microlensing, and $n$ the number of detected microlensing events. Using the published MACHO and OGLE data one finds that $N\tau/n \approx 50 - 100$. A major improvement should be possible if various sources of noise in the data could be better controlled.

The number of stars that can be monitored from the ground is inversely proportional to the square of seeing disk diameter, as the detection limit is set by crowding, not by photon statistics. This implies that the sites with the best possible seeing combined with small, 1-meter class telescopes and the largest possible number of pixels offer the best price to performance ratios. I am not aware of a sound estimate of the number of stars measurable from the ground, but it may be as high as $10^9$, and over $\sim 10^3$ microlensing events per year may be within reach with a modest extension of the current technology.

## 2. Variable stars

The massive photometric searches for gravitational microlensing events lead to the discovery of a huge number of variable stars (Udalski et al. 1994e, 1995, Grison et al. 1995, Alcock et al. 1995b, Cook et al. 1995, Kaluzny et al 1995a,b). The total number of variable stars in the data bases of the four collaborations is $\sim 10^5$, with MACHO having by far the largest number of unpublished objects (Alcock 1995b, Cook et al. 1995), and OGLE having the largest number of published finding charts, light curves, etc. (Udalski et al. 1994e,1995).

I would like to concentrate on just one type of variables: the detached eclipsing binaries, which are also double line spectroscopic binaries, as these are the primary source of the fundamental data about stars: their masses, luminosities, and radii (Andersen 1991, and references therein). Such systems are relatively rare, roughly one out of a few thousand. As the eclipses are narrow it is necessary to accumulate a few hundred photometric measurements to determine the binary period. This means that one needs a total of $\sim 10^6$ photometric measurements to discover one promising candidate for a detached eclipsing binary, which is also likely to be a double line spectroscopic binary. Massive photometric programs are needed, and this is exactly what is provided by the microlensing searches. The light curves and finding charts for many such objects are provided by Udalski et al. (1994a, 1995) for stars towards the galactic bulge, and by Grison et al. (1995) for stars in the LMC.

Theoretical analysis of the light variations during the two eclipses of a detached binary gives the (star size) / (orbit size) ratio for the two stars, $R_1/A$, and $R_2/A$, as well as the fraction of total luminosity (in a given photometric band) originating in each component, $L_1/L$, and $L_2/L$. The so called "third light", defined as: $L_3 \equiv L - L_1 - L_2 \neq 0$, may also be present and readily measured (Dreshel et al. 1989, Goecking et al. 1994, Gatewood et al. 1995, and references therein). Finally, the inclination of the orbit $i$ can be determined.

If the two stars have similar luminosities then the spectra of both stars, and the amplitudes of radial velocity variations of both stars, $K_1$ and $K_2$, can be measured. Given $K_1$ and $K_2$, as well as the orbital period and inclination, the size of the binary orbit, $A$, and the two masses, $M_1$ and $M_2$, can be determined. With $A$ as well as $R_1/A$ and $R_2/A$ known, the two stellar radii, $R_1$ and $R_2$, become known as well. The practical application of this procedure to real bright stars lead to the determination of all stellar parameters with a $\sim 1\% - 2\%$ accuracy for a few dozen objects with spectral types all the way from O8 to M1 (Andersen 1991).

## 3. Detached eclipsing binaries as primary distance indicators

The fact that all parameters for the bright binaries can be obtained with $\sim 1 - 2\%$ accuracy implies that it should be possible to measure distances to far away binaries with $\sim 1 - 2\%$ accuracy, provided very high S/N data can be obtained with the large telescopes. There are elements which can make this task difficult to accomplish in practice: unresolved companions and interstellar extinction. There is reason for optimism, as the problem of unresolved companions, i.e. the "third light" problem, is well known among the brightest objects, like Algol (Gatewood et al. 1995), and given high enough S/N data the "third light" contribution can be calculated. The interstellar extinction is a standard problem for any photometric distance determination, and it is always troublesome. It helps to have infrared photometry, as the interstellar extinction in the K-band is several orders of magnitude lower than it is in the visual domain.

In order to measure the distance we have to find out what is the surface brightness of each binary component in our photometric bands. This is the only delicate step in the whole procedure. The surface brightness can be obtained from the observed colors and/or spectra using modern model atmospheres. This relation can be well calibrated empirically with the nearby systems, which have their distances accurately measured either with trigonometric parallaxes (Dommanget & Lampens 1992, Monet et al. 1992, Gatewood 1995, Gatewood et al. 1995), or with a combination of spectroscopic and interferometric (astrometric) orbits (Pan et al. 1990, 1992). The most accurate surface brightness determination is possible in the K-band (Ramseyer 1994).

Given a good estimate of the surface brightness of each star in a selected band, say $F_K^*$, as well as the direct measurement of the flux in that band at the telescope, $F_K$, we can calculate the distance as

$$d = R \times \left(\frac{F_K^*}{F_K}\right)^{1/2},$$

where $R$ is the stellar radius, and $F_K$ has been corrected for the interstellar extinction.

This approach was attempted by Bell et al. (1991, 1993) to measure the distance to LMC. Unfortunately, the binaries HV 2226 and HV 5963 had light curves indicating these were semi-detached, i.e. moderately complicated system. Also, the accuracy of radial velocity measurements was rather low. The list of EROS binaries in LMC (Grison et al. 1995) has a number of clearly detached systems, as judged from their light curves. Very accurate measurements of radial velocities of both components of a binary system are now possible with the recently developed TODCOR method (Zucker & Mazeh, 1994). Metcalfe et al. (1995) used TODCOR to measure radial velocity amplitudes for a 13 mag eclipsing binary CM Dra with a precision of 0.15 $km\ s^{-1}$ with a 1.3 meter telescope.

The detached binaries offer a potential to establish distance to globular clusters, to the galactic center, and to nearby galaxies: LMC, SMC, M31 and M33, with unprecedented accuracy. Note, that binaries cover a very the wide range of spectral types, and they are numerous, so cross-checks will be possible. If the crowding and interstellar extinction can be handled adequately, the fractional accuracy of distances will be equal to the fractional accuracy of the determination of radial velocity amplitudes, $K_1$ and $K_2$, as distances are proportional to stellar diameters, which in turn are proportional to the $K$ values. Eclipsing binaries offer a "single step" distance determination to nearby galaxies, thereby providing an accurate zero point calibration for all pulsating stars, including cepheids – a major step towards very accurate determination of the Hubble constant.

## 4. Detached eclipsing binaries as primary age indicators

The accurate age determination of globular clusters is one of the most important astronomical issues, as there is a perceived conflict with the determinations of the Hubble constant (Chaboyer et al. 1995). All current age determinations are based on the comparison between the observed and theoretical isochrones in the color-magnitude diagrams (Shi 1995, and references therein). There are at least two problems with this method. First, the age is inversely proportional to the square of distances to a globular clusters, and second, the colors of theoretical models are affected by poorly known and not understood "mixing-length" parameter (Paczyński 1984, Chaboyer 1995, and references therein). Recent discovery of the two detached eclipsing binaries at the main sequence turn-off point in Omega Centauri (Kaluzny et al. 1995b) is a major step to overcome both problems.

The detached binaries in Omega Centauri will be used to accurately measure the distance to this globular cluster with the method described in the previous section. This will considerably reduce the distance uncertainty and the corresponding uncertainty in the classical age determination. Also, for the first time ever it will be possible to measure directly the masses of stars near the main sequence turn-off point, and this in turn will allow the age determination using the mass-luminosity relation, which is not affected by the "mixing-length" parameter (Paczyński 1984). The masses of stars near the turn-off point depend on the age as well as helium content Y, and heavy element content Z. The Z abundance can be measured spectroscopically, but the helium abundance has to be determined from the same mass-luminosity relation as the age. To make this possible it will be necessary to discover detached eclipsing binaries somewhat below the turn-off point, and to measure their masses.

## 5. Microlensing by Internet

The photometry of OGLE microlensing events, their finding charts, as well as a regularly updated OGLE status report, including more information about the "early warning system", can be found over Internet from the host: "sirius.astrouw.edu.pl" (148.81.8.1), using "anonymous ftp" service (directory "ogle", files "README", "ogle.status", "early.warning"). The file "ogle.status" contains the latest news and references to all OGLE related papers, and PostScript files of some publications. These OGLE results are also available over World Wide Web at: "http://www.astrouw.edu.pl".

Similar information about MACHO results is available over World Wide Web at: "http://darkstar.astro.washington.edu".

## 6. Acknowledgemens

It is a great pleasure to acknowledge the discussions with, and comments by Dr. Dr. J. Kaluzny, A. Kruszewski, D. W. Latham, M. Pratt, and C. Stubbs. This work was supported by the NSF grants AST-9216494 and AST-9313620.